\documentclass{aa}
\usepackage{amsmath}
\usepackage{amsfonts}
\usepackage{todo}
\usepackage{upgreek}
\usepackage[T1]{fontenc}
\usepackage[utf8]{inputenc}
\usepackage{lmodern}
\usepackage{ulem}
\usepackage{float}

\graphicspath{ {files/} }

\begin{document}
\title{Kernel nullers for an arbitrary number of apertures}
\author{Romain Laugier \inst{1}
        \and
        Nick Cvetojevic \inst{1}
        \and
        Frantz Martinache \inst{1}
        }
\institute{
        Université Côte d'Azur, Observatoire de la Côte d'Azur, CNRS, Laboratoire Lagrange, France
        }

\abstract
{The use of interferometric nulling for the direct detection of extrasolar planets is in part limited by the extreme sensitivity of the instrumental response to tiny optical path differences between apertures. The recently proposed kernel-nuller architecture attempts to alleviate this effect with an all-in-one combiner design that enables the production of observables inherently robust to residual optical path differences ($\ll \lambda$).
}
{Until now, a unique kernel nuller design has been proposed ad hoc for a four-beam combiner. We examine the properties of this original design and generalize them for an arbitrary number of apertures.
}
{We introduce a convenient graphical representation of the complex combiner matrices that model the kernel nuller and highlight the symmetry properties that enable the formation of kernel nulls. The analytical description of the nulled outputs we provide demonstrates the properties of a kernel nuller. 
}
{Our description helps outline a systematic way to build a kernel nuller for an arbitrary number of apertures. The designs for three- and six-input combiners are presented along with the original four-input concept. The combiner grows in complexity with the square of the number of apertures. While one can mitigate this complexity by multiplexing nullers working independently over a smaller number of sub-apertures, an all-in-one kernel nuller recombining a large number of apertures appears as the most efficient way to characterize a high-contrast complex astrophysical scene.
}
{One can design kernel nullers for an arbitrary number of apertures that produce observable quantities robust to residual perturbations. The designs we recommend are lossless and take full advantage of all the available interferometric baselines. They are complete, result in as many kernel nulls as the theoretically expected number of closure-phases, and are optimized to require as few outputs as possible.
}

\maketitle

\section{Introduction}
    The last 25 years have seen the detection of more than 4,000 exoplanets \citep{Schneider2011}. Despite the indirect nature of most detections, existing observations already provide us with a wealth of information on the properties of exoplanetary systems: their mass, size, and orbital elements. Yet direct detection of a planet's reflected, or radiated light, for its direct spectral analysis for a large sample of targets, remains an exciting prospect that will contribute to further characterize individual planets, in particular the properties of their atmospheres \citep{Marois2008, Zurlo2015}.\par
    
    The use of coronagraphic instruments is now leading to the detection of young giant planets in wide orbits around nearby stars \citep{Macintosh2015, Chauvin2017, Mesa2019}. This success is carried by the continued improvements of extreme adaptive optics systems \citep{Sauvage2016, Lozi2018, Boccaletti2020}. For smaller separations approaching the diffraction limit (below $\sim3\lambda/D$), small residual wavefront errors still dominate the error budget and coronagraphic solutions become less favorable.\par
    
    \cite{Lacour2019} have demonstrated the advantages brought by long baseline interferometry for the characterization of extrasolar planets. This observing mode takes advantage of the spatial filtering provided by the resolving power of each of the 8 meter telescopes of the VLTI, coupled into single mode fibers to reach the required contrast. Interferometric nullers \citep{Bracewell1978, Colavita2009, Serabyn2019, Hoffmann2014, Defrere2015, Norris2019} offer the possibility to explore smaller angular separations through the use of fragmented apertures and long-baseline interferometry. Some combining solutions have been found that optimize the rejection of resolved stars \citep{Angel1997, Guyon2013}. The exploitation of these instruments is still limited by their vulnerability to optical path differences (OPD) errors, and requires sophisticated statistical analysis, like those proposed by \cite{Hanot2011}, \cite{Defrere2016} and more recently used by \cite{Norris2019} to disentangle the off-axis astrophysical signal from the effects of unwanted OPD.\par
    
    Classical long-baseline and Fizeau interferometry make extensive use of the production of robust observables, like closure phases \citep{Jennison1958} and their generalized form, kernel phases \citep{Martinache2010}, to sidestep the limitations brought by the OPD residuals. This approach has provided reliable performance at very small separations, down to one resolution element and below.\par
    
    Bringing together the robustness of interferometric observables and the photon-noise suppression of nulling, is an exciting perspective as it opens a novel high-contrast, high-precision regime. The double Bracewell architecture \citep{Angel1997} was remarked to offer such robustness \citep{Velusamy2003} when implemented with the adequate phase shift between the two stages (called "sin-chop"). A different approach was later proposed by \cite{Lacour2014}, exploiting the measurement of fringes in the leakage light.\par
    
    \cite{Martinache2018} introduced an alternative, more efficient solution for a four-telescope beam-combiner architecture that produces six nulled outputs. By analyzing the response of these outputs to parasitic OPDs (instrumentally or atmospherically induced phase error), the authors identify linear combinations of outputs that are robust to these aberrations to second order. The solution they propose concerns a four-input nuller that provides three nulled kernel observables.\par
    
    In this paper, we look for the properties ensuring that a combiner will produce kernel nulls. They help outline a general strategy for the design of kernel nullers for an arbitrary number of apertures.\par

\section{Analysis of the existing four-input kernel-nulling architecture}
  
  \subsection{The kernel-nulling approach} \label{sec_kernel_principle}
    Using the nuller architecture laid out by \cite{Martinache2018} for a four-beam interferometer as a starting point, and reexamining its properties, we look into ways of generalizing this special case to a wider range of configurations, involving different numbers of apertures.\par
    
    The inner structure of a homodyne interferometric combiner (nulling or not) is conveniently represented by a combiner matrix $\mathbf{M}$ that acts on a vector $\mathbf{z}$ of input electric fields and leads to the production of an output electric field vector $\mathbf{x}$.
    \begin{equation}\label{eq_xMz}
        \mathbf{x} = \mathbf{M} \cdot \mathbf{z}.
    \end{equation}
    Eventually, a detector records the intensity associated to the square norm of this output electric field.\par
    
    The fact that only the square norm of the field is recorded has two consequences. The first is that the response of the combiner is insensitive to the absolute phase of the input electric field: one of the sub-apertures can therefore be arbitrarily picked as a reference, and the phases of the different electric fields sampled by the other sub-apertures are measured relative to that reference. The second is that the output intensity is equally insensitive to any global phase shift $\phi$ applied to any row $\mathbf{m}$ of the matrix $\mathbf{M}$ describing the combiner. This is of consequence when identifying distinct combinations (or rows).\par
    
    Assuming that the recombiner is fed by a balanced array of identical sub-apertures, the complex amplitude of the input electric field can be described by a vector of phasors. We will further assume that the combiner benefits from a fringe tracker that, although not perfect, brings the system close to its nominal state. The fringe tracking residuals are assumed to be small and all phasors $e^{-j\varphi_k}$ are here approximated using the following expansion:
    \begin{equation}\label{eq_simplified_inputs}
        z_{k} = e^{-j \varphi_k} \approx 1 -j \varphi_k,
    \end{equation}
    where $j$ is the imaginary unit.\par
    
    Since only intensities are measured, the overall response of the system is a quadratic function of the perturbation phase vector. \cite{Martinache2018} thus use this description to look at the properties of the second order derivative of the intensity relative to the phase. One of the $n_a$ sub-apertures being used as a phase reference, there are $n_a - 1$ degrees of freedom, and $n_d = n_a * (n_a - 1) / 2$ such derivatives. One can store this response into a $n_o \times n_d$ matrix $\mathbf{A}$ called the matrix of second order derivatives, where $n_o$ is the number of relevant outputs. Linear combinations of rows of $\mathbf{A}$ that equal 0 cancel out the second order intensity deviations caused by small input phase errors. The same linear combination applied to the intensity measured after the recombiner will be equally insensitive to small input phase errors. We refer to these linear combinations as kernel outputs or kernel nulls when applied to a collection of nulled outputs.\par
    
    The rank of $\mathbf{A}$ and the possibility of forming such robust observables rely entirely on the properties of the matrix $\mathbf{M}$, and therefore does not depend on the geometry of the input array. However, the question of whether a kernel null carries astrophysically relevant information also depends on the configuration of the array. Throughout this work, phase and amplitude contributions are considered independently, but their coupled contribution is neglected. For now, we will further examine the properties of the combiner and introduce a convenient visual representation of the structure of $\mathbf{M}$.

    \subsection{Visualization of complex combiner matrices}\label{sec_complex_matrix_plots}
        The effect of the matrix $\mathbf{M}$ on the complex amplitude of the input electric field can be conveniently visualized by a series of plots of the complex plane.  For a given combiner, each input is represented by a colored arrow which, in the absence of environmental perturbation, is aligned with the real axis. Each plot illustrates the effect of a row of $\mathbf{M}$ on such inputs: the resulting electric field is the sum of all colored arrows present in the plot. A nuller is characterized by several outputs for which the sum of the arrows, associated to the electric fields, sum up to zero. These complex matrix plots (CMP) will be used throughout this work to describe several nuller designs of varying complexity.\par

    \subsection{From real to complex nulls} \label{sec_real_to_complex}
\begin{figure}
            \centering
            \includegraphics[width=0.49\textwidth]{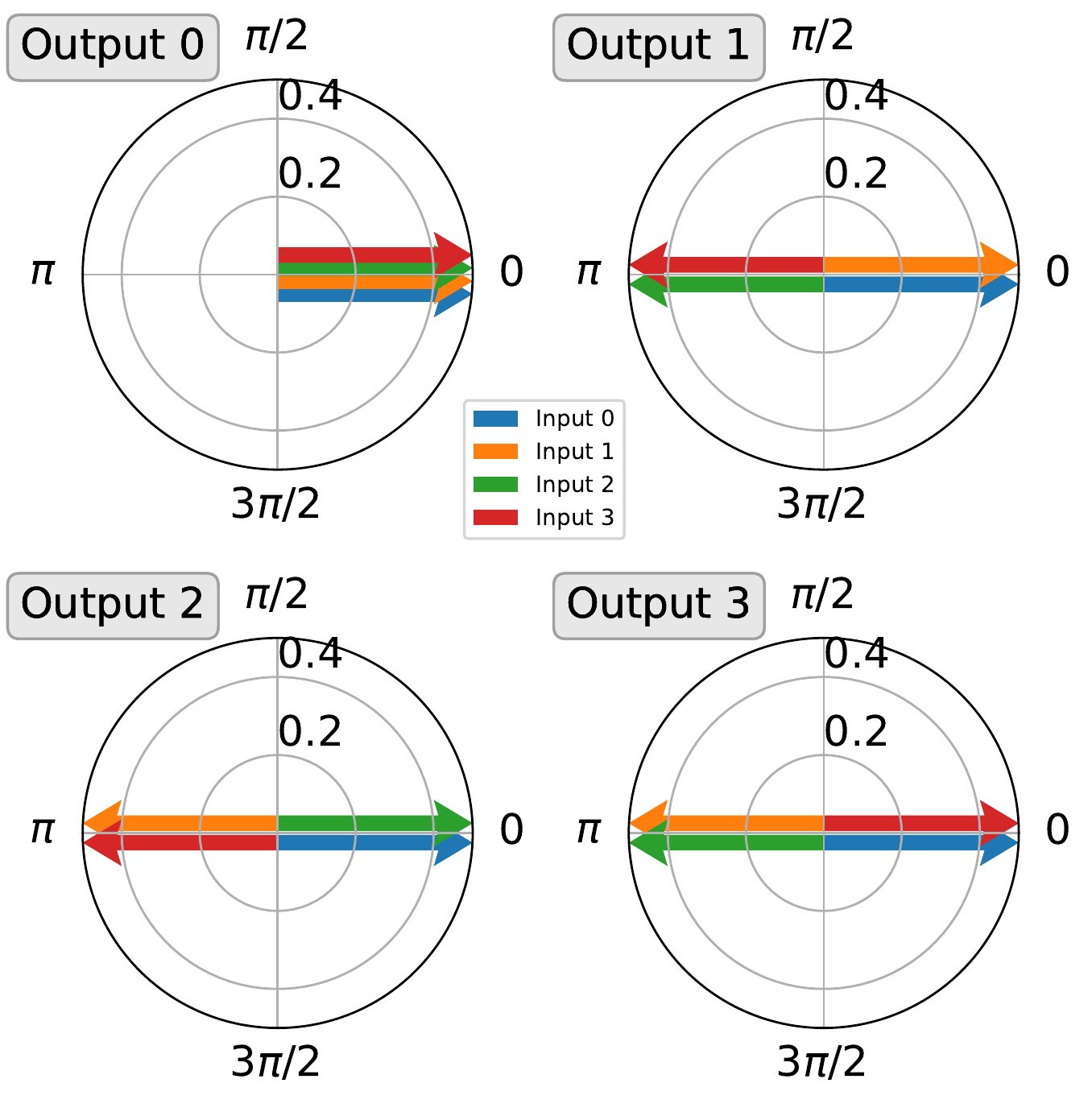}%{nick2.pdf} %
            \caption{CMP of the matrix $\mathbf{N}_4$ of Eq. (\ref{eq_N}) representing a four-input nulling combiner. The first row constitutes the bright channel, with all inputs combined constructively. Note how each output is a contribution of all the inputs, and not just a pair of them, which prevents the direct interpretation through the uv plane.}
            \label{fig_paper_plot_N}
        \end{figure}

\begin{figure}
            \centering
            \includegraphics[width=0.4\textwidth]{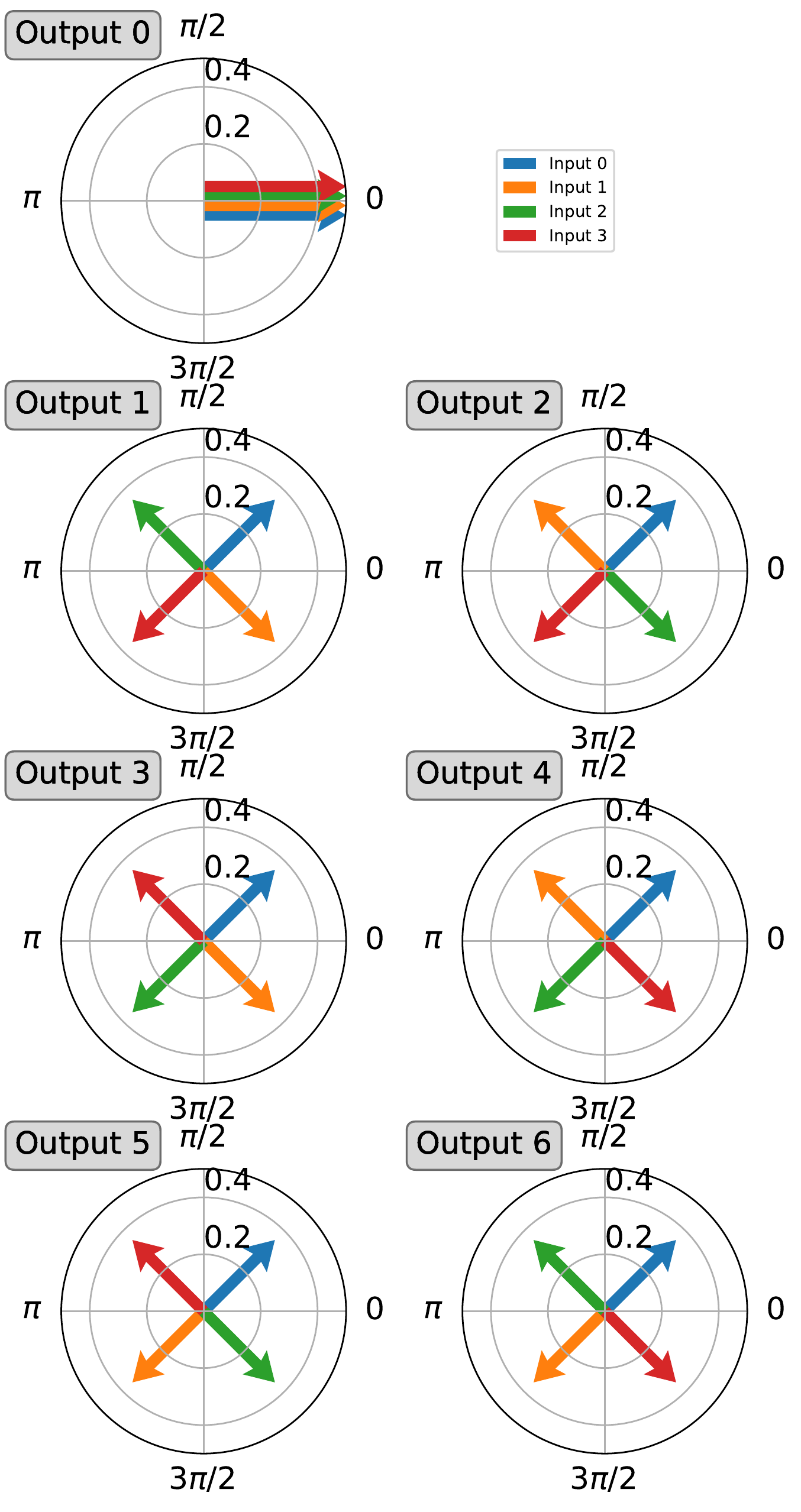}%{nick3b.pdf}
            \caption{CMP of the $\mathbf{S}_4\cdot\mathbf{N}_4$ combination. The first output is the bright channel for which all the inputs add-up constructively. The vectors are staggered for readability. Pairs of nulled rows represented side-by-side are mirror images of each-other (enantiomorph). Note that the amplitude of the phasors is reduced compared to Fig. \ref{fig_paper_plot_N} due to additional splitting.}
            \label{fig_paper_plots}
        \end{figure}

        The architecture of the kernel nuller described in \cite{Martinache2018} builds from an initial all-in-one four-beam nuller whose overall effect can be described by the following matrix:
        \begin{equation}\label{eq_N}
            \mathbf{N}_4 =
                \frac{1}{\sqrt{4}}\left[\begin{matrix}
                    1 &  1 &  1 &  1\\
                    1 &  1 & -1 & -1\\
                    1 & -1 &  1 & -1\\
                    1 & -1 & -1 &  1\\
                            \end{matrix}\right]
        \end{equation}
        This matrix is real. Each nulled row of $\mathbf{N_4}$ recombines distinct arrangements of the four input electric fields such that the coefficients on the corresponding rows sum up to zero, as represented in Fig. \ref{fig_paper_plot_N}, with arrows aligned with the real axis: two positive (or not phase-shifted), and two negative (or phase-shifted by $\pi$). As discussed in this reference, this nuller does not allow the formation of kernels: the output intensities it produces are a degenerate function of the target information and input phase perturbations. The outputs of this nuller can however be fed to a second stage, described by the following matrix:
        \begin{equation}\label{eq_S}
            \mathbf{S}_4 =
                \frac{1}{\sqrt{4}}\left[\begin{matrix}
                    2 &  0  &  0 & 0 \\
                    0  & 1 & e^{j\frac{\pi}{2}} &  0\\
                    0  & e^{j\frac{\pi}{2}} & 1 & 0\\
                    0  & 1 & 0 & e^{j\frac{\pi}{2}}\\
                    0  & e^{j\frac{\pi}{2}} & 0 & 1\\
                    0  & 0 & 1 & e^{j\frac{\pi}{2}}\\
                    0  & 0 & e^{j\frac{\pi}{2}} & 1\\
                            \end{matrix}\right],
        \end{equation}
        which leaves the bright output untouched but further splits the nulled ones, and selectively introduces $\pi/2$ phase shifts. The overall effect of the combiner will be described by a now complex combiner matrix, result of the product $\mathbf{M}_4 = \mathbf{S}_4 \cdot \mathbf{N}_4$. The CMPs of this modified combiner, shown in Fig. \ref{fig_paper_plots}, offer a more easily readable description of its effect, with components of the output electric field no longer simply aligned with the real axis, but spanning the complex plane.\par
        
        The new complex configuration enables the larger diversity that is required to disentangle the otherwise degenerate effect that environmental perturbations have on the input electric fields. The modified nuller indeed features more outputs than inputs, and a close examination of the CMPs of Fig. \ref{fig_paper_plots} shows that all six combinations offer a distinct arrangement of the four input fields. The construction of a larger number of distinct nulls is one of the requirements for the existence of a non-empty left null space for $\mathbf{A}$ described in Sec. \ref{sec_kernel_principle}. In effect, pairs of outputs produce the same response to environmental effects, while still producing different response to off-axis light.\par

\section{Properties of conjugate pairs of nulls}\label{sec_properties}
    \subsection{Kernel outputs}
        Identifying the kernel-forming combinations of outputs no-longer requires building the second order derivative matrix $\mathbf{A}$ but can be achieved by examination of the CMP representation of the nuller.   Figure \ref{fig_paper_plots} lays out, side by side, the two outputs leading to one kernel-null observable. We call these outputs enantiomorph: close examination of any such pair of outputs reveals that the electric field combination patterns are the mirror image of one-another.\par
        
        Given that the measured intensity associated to any output is insensitive to a global phase shift, one can always apply such a shift so as to align the arrow corresponding to the phase reference input with the real axis, and point it towards the positive direction. After such a rotation is applied, enantiomorph outputs simply become complex conjugate. This makes it possible to write simple equations that describe the two key properties of kernel nulls: their robustness to small phase perturbation, and the antisymmetric nature of the signal they provide.

    \subsection{Robustness} \label{sec_robustness}
        
        Considering $\mathbf{m}_1$ and $\mathbf{m}_2$, a conjugate pair of null rows of $\mathbf{M}$:
        \begin{equation}\label{eq_m}
            \mathbf{m}_2 = \mathbf{m}_1^*.
        \end{equation}
        A corresponding kernel null $\kappa(\mathbf{z})$ writes as the difference of the two measured intensities:
        \begin{equation}\label{eq_km}
            \kappa(\mathbf{z}) = 
            |\mathbf{m}_1 \mathbf{z}|^2 - 
            |\mathbf{m}_2 \mathbf{z}|^2 =
            \mathbf{m}_1 \mathbf{z} (\mathbf{m}_1 \mathbf{z})^* - 
            \mathbf{m}_2 \mathbf{z} (\mathbf{m}_2 \mathbf{z})^*.
        \end{equation}
        Using (\ref{eq_m}) and (\ref{eq_km}) gives:
        \begin{equation}\label{eq_mstar}
            \kappa(\mathbf{z}) = 
            \mathbf{m}_1 \mathbf{z} (\mathbf{m}_1 \mathbf{z})^* - 
            \mathbf{m}_1^* \mathbf{z} (\mathbf{m}_1^* \mathbf{z})^*.
        \end{equation}
        In the case of the approximation mentioned in Eq. (\ref{eq_simplified_inputs}):
        %\begin{equation}
            \begin{multline}
                \kappa(\mathbf{z}) =
                \mathbf{m}_1 (\mathbf{a} + j\boldsymbol{\varphi}) (\mathbf{m}_1 (\mathbf{a} + j\boldsymbol{\varphi}))^*\\ 
                - \mathbf{m}_1^* (\mathbf{a} + j\boldsymbol{\varphi}) (\mathbf{m}_1^* (\mathbf{a} + j\boldsymbol{\varphi}))^*,
            \end{multline}
        %\end{equation}
        where $\mathbf{a}$ is a vector of ones. Developing this expression, since $\mathbf{m}_1\mathbf{a} = 0$ and $\mathbf{m}_1^*\mathbf{a} = 0$, the only terms left are the ones containing only the imaginary perturbation term $j\boldsymbol{\varphi}$:
        \begin{equation}
            \kappa(\mathbf{z}) = 
            \mathbf{m}_1 j\boldsymbol{\varphi} (\mathbf{m}_1 j\boldsymbol{\varphi})^* - 
            \mathbf{m}_1^* j\boldsymbol{\varphi} (\mathbf{m}_1^* j\boldsymbol{\varphi})^*.
        \end{equation}
        Distributing the conjugate operator gives:
        \begin{equation}
            \kappa(\mathbf{z}) = 
            - \mathbf{m}_1 j\boldsymbol{\varphi} \mathbf{m}_1^* j\boldsymbol{\varphi} + 
            \mathbf{m}_1^* j\boldsymbol{\varphi} \mathbf{m}_1 j\boldsymbol{\varphi},
        \end{equation}
        and therefore $\kappa(\mathbf{z}) = 0$ due to the commutativity. This shows that the subtraction of intensity of complex conjugate pairs of nulled outputs always produces a kernel null that is robust to arbitrary imaginary phasors, to which the small input phase aberrations are approximated.\par
        
        This property also applies to arbitrary purely real input electric fields that would correspond to pure photometric error generated by fluctuations of the coupling efficiencies. Considering a purely real input vector $\mathbf{a}$:
        \begin{equation}
            \kappa(\mathbf{z}) = 
            \mathbf{m}_1 \mathbf{a} (\mathbf{m}_1 \mathbf{a})^* - 
            \mathbf{m}_1^* \mathbf{a} (\mathbf{m}_1^* \mathbf{a})^*.
        \end{equation}
        Distributing the conjugate operator gives:
        \begin{equation}\label{eq_robust_last}
            \kappa(\mathbf{z}) = 
            \mathbf{m}_1 \mathbf{a} \mathbf{m}_1^* \mathbf{a} - 
            \mathbf{m}_1^* \mathbf{a} \mathbf{m}_1 \mathbf{a},
        \end{equation}
        and therefore $\kappa(\mathbf{z}) = 0$ due to the commutativity.\par
        
        At any instant, the subtraction of the signals recorded by conjugate (or more generally enantiomorph) outputs forms a kernel null.
        Conjugate pairs of nulls allow the formation of kernel nulls. This property generalizes to enantiomorph pairs of nulls through the rotation by a single common phasor. A complementary approach for the identification of robust combinations of outputs is the use of the singular value decomposition (SVD) of the second order derivative matrix $\mathbf{A}$, as mentioned by \cite{Martinache2018}, which ensures the identification of all the robust combinations of outputs.\par
        
\begin{figure*}
            \centering
            \begin{tabular}{c|c|c}
                Small error & Partially resolved & Null peak\\
                \includegraphics[width=0.31\textwidth]{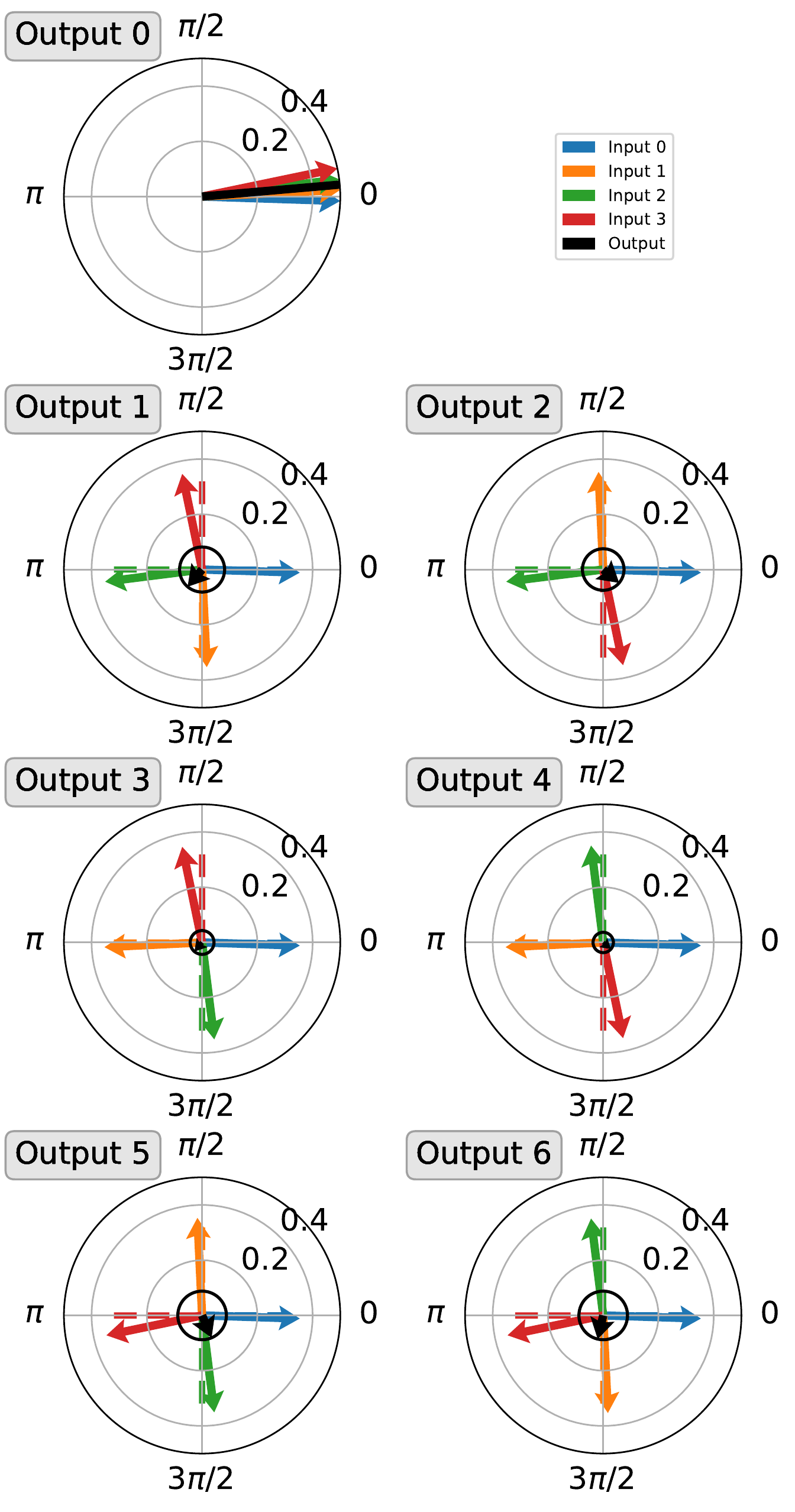}&
                \includegraphics[width=0.31\textwidth]{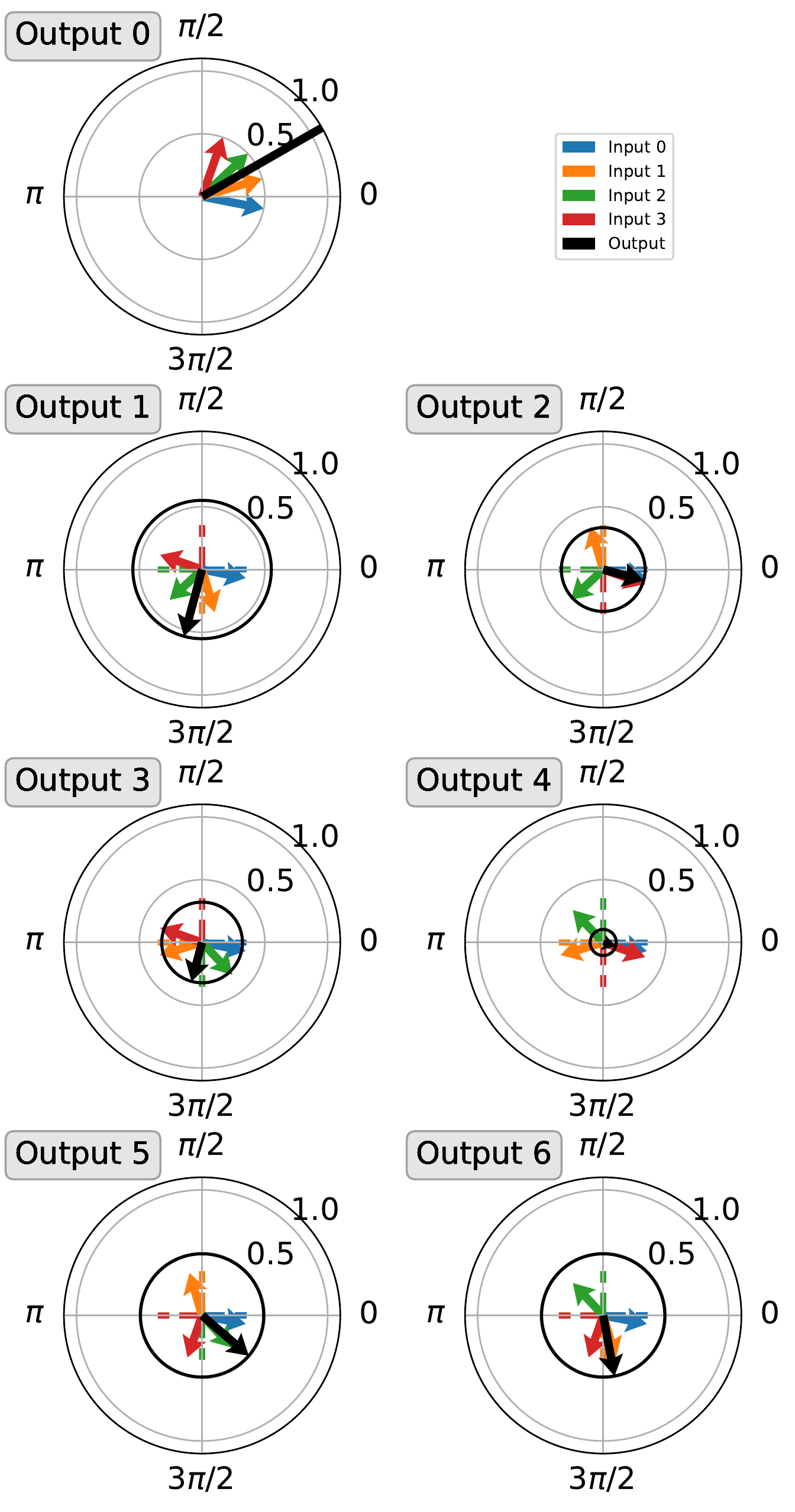}&
                \includegraphics[width=0.31\textwidth]{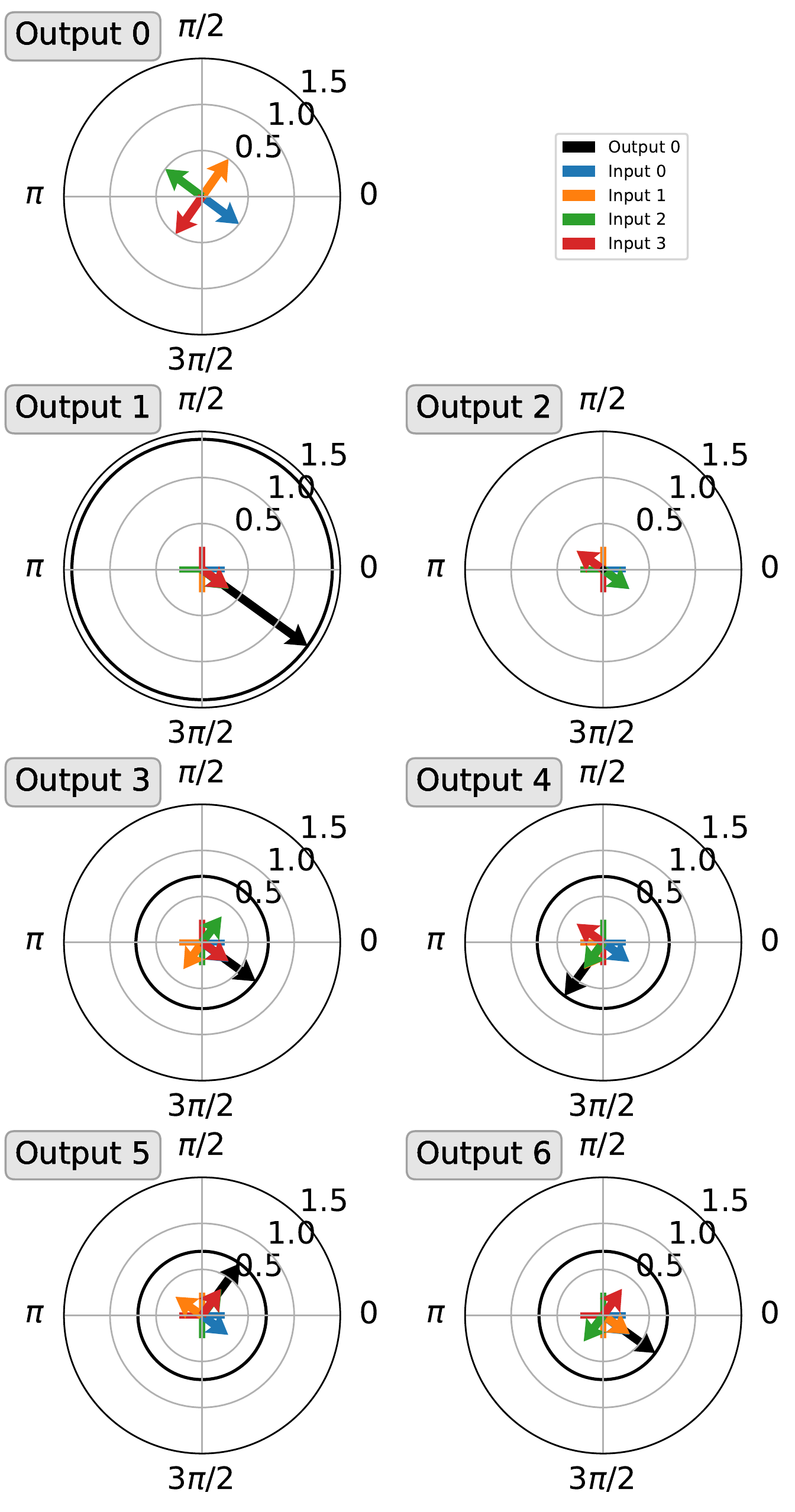}\\
            \end{tabular}
            \caption{CMP for a four-input combiner, representing in dashed lines the coefficients of the combiner matrix $\mathbf{M}_4'$, functionally equivalent to $\mathbf{M}_4$, and in solid arrows the contributions of the complex amplitude of an example input electric field to the output electric field represented in black. Most dashed lines are hidden under the arrows. A black circle of radius equal to the modulus of this output is plotted for visual cue, its area being proportional to the corresponding intensity. Enantiomorph pairs that generate kernel combinations by subtraction are represented side-by-side. Like \cite{Martinache2018}, we use the example of the VLTI UT configuration observing at zenith at a wavelength of $3.6\mu m$. On the left-hand panel, a source located at 0.2 mas from the optical axis and for which the corresponding input phase shifts are within the small phase approximation. As a result, the output intensities within each pair are fully correlated and result in no kernel signal. On the central panel, the source is located 1.1 mas off-axis which generates larger phase shifts. As a result, the null intensities from the enantiomorph pairs begin to decorrelate and generate kernel-null signal. On the right-hand panel, the source is located 4.3 mas from the optical axis, in the position where the first nulled output peaks. At this position, the second output gets to zero.}
            \label{fig_didactic_plots}
        \end{figure*}

        This behavior can be illustrated by adding different phased contributions to the inputs, and plotting the resulting electric field on top of the original perfectly cophased CMP (kept in dashed lines). The first panel of Fig. \ref{fig_didactic_plots} uses this representation of the combined light to illustrate how small input phase aberrations affect the amplitude (and therefore the intensity) of the combiner's outputs. In particular it shows how, for small phase errors,  conjugate pairs of nulls suffer the same leakage light intensity.\par

    \subsection{Symmetry of the response}\label{sec_odd_response}
        The second and third panels of Fig. \ref{fig_didactic_plots} show how input light coming from a significantly off-axis source (input phases $\phi \geq 1$ radian) propagates to the nulled outputs. They highlight how this off-axis light produces different intensities at the outputs of the conjugate pairs, translating into kernel-null signal.\par
        
        For a combiner that is fed by an array of apertures collecting the light from the sky, the value of this response as a function of the incidence of the light is the response map of the interferometer, and depends on the position of each of the apertures. \cite{Martinache2018} note how this map is antisymmetric, therefore providing a rejection of the photosphere of stars and symmetric circumstellar disks that could hide a planetary companion, and provide the astrometry of such companions without ambiguity. Using our formalism, we can demonstrate this antisymmetric property for any aperture configuration. If $\mathbf{z}$ and $\mathbf{z}'$ are two input electric field vectors coming from sources located at symmetric positions in the field of view, then:
        \begin{equation}\label{eq_z}
            \mathbf{z}' = \mathbf{z}^*.
        \end{equation}
        Considering again a conjugate pair of null rows $\mathbf{m}_1$ and $\mathbf{m}_2$, and by substitution of (\ref{eq_z}) into (\ref{eq_km}) we get:
        \begin{equation}
            \kappa(\mathbf{z}) = 
            \mathbf{m}_1 \mathbf{z}'^* \mathbf{m}_1^* \mathbf{z}' - 
            \mathbf{m}_2 \mathbf{z}'^* \mathbf{m}_2^* \mathbf{z}'.
        \end{equation}
        After substitution of (\ref{eq_m}), this becomes:
        \begin{equation}
            \kappa(\mathbf{z}) = 
            \mathbf{m}_2^* \mathbf{z}'^* \mathbf{m}_2 \mathbf{z}' - 
            \mathbf{m}_1^* \mathbf{z}'^* \mathbf{m}_1 \mathbf{z}'.
        \end{equation}
        This leads to the conclusion that the response is antisymmetric:
        \begin{equation}\label{eq_response_last}
            \kappa(\mathbf{z}) = - \kappa(\mathbf{z}').
        \end{equation}
        \par

        Conversely, one may also extract from this pair of nulls the complementary observable:
        \begin{equation}\label{eq_tau}
            \tau(\mathbf{z}) =
            \mathbf{m}_1 \mathbf{z} (\mathbf{m}_1 \mathbf{z})^* +
            \mathbf{m}_2 \mathbf{z} (\mathbf{m}_2 \mathbf{z})^*
        \end{equation}
        whose response is symmetric. The observables $\kappa$ and $\tau$ therefore carry complementary information on the target field, much like the amplitude and phase of complex visibility in classical interferometry.\par
        
        Although $\tau$ does not have the same robustness to aberrations, there may be ways to use it with the processing methods employed by \cite{Hanot2011} and \cite{Norris2019} so as to provide additional information on the target in different science cases. $\kappa$ is best suited for the study of high-contrast non-symmetrical features such as planetary companions, while $\tau$ may be used to study brighter symmetrical features such as debris disks or stellar envelopes. Combining both types of observables could enable image reconstruction.\par
        
        The $\tau$ observables carry some information about the input phase errors. One can use their values over the course of a scan or modulation of the OPDs to locate the setpoint of the kernel-nuller, for which they will reach a minimum.

\section{Construction of new nullers}\label{sec_construction}
        
  \subsection{Blueprints of kernel-nulling matrices} \label{sec_generative}
        The properties used in Sects. \ref{sec_robustness} and \ref{sec_odd_response} to demonstrate the robustness of kernel nullers to small phase perturbations may be used as constraints to guide the design of an arbitrary kernel nuller matrix. For the output of any row $l$ to provide an on-axis null, the matrix coefficients must satisfy:
        \begin{equation}\label{eq_null_condition}
            \sum_{k=0}^{n_a-1}{M_{k,l}} = 0.
        \end{equation}
        Output intensities are unchanged when the coefficients of a row are all multiplied by a common phasor. We therefore apply one such phasor so as to get $\mathrm{Arg}(M_{0,l}) = 0$. We also set output \#0 by combining all the inputs with zero phase offset: $\mathrm{Arg}(M_{k,0}) = 0$.\par

        Simple solutions to Eq. (\ref{eq_null_condition}) for a balanced array can be found by picking arrangements of uniformly spaced phase values in the $[0,2\pi]$ interval as can be seen of Figs.  \ref{fig_paper_plots}, \ref{fig_3x3} and \ref{fig_6T_combiner}. The phase of each coefficient is therefore a multiple of $\Phi_0 = 2\pi/n_a$. On the CMPs seen thus far, this would result in the rotation of all of the arrows on the nulled outputs until the one associated with input \#0 is aligned with the real axis in the positive direction. With these constraints in place, outputs will only differ in the order in which the remaining $n_a - 1$ phase offsets are associated with the inputs. The maximum number of distinct nulled outputs is therefore:
        \begin{equation}\label{eq_nmax}
            n_{max} = (n_a-1)! .
        \end{equation}
        \par

        The phase term $\phi_{k,l}$ writes:
        \begin{equation}
            \phi_{k,l} = c_{k,l} \Phi_0,
        \end{equation}
        where $c_{k,l}a$ is the $k$-th term of the $l$-th possible combination on the circle. In general, a complex coefficient of $\mathbf{M}$ will therefore write:
        \begin{equation}\label{eq_generative}
            M_{k,l} = a_l \cdot e^{j \phi_{k,l}},
        \end{equation}{}
        where $a_l$ is a real coefficient, normalizing the matrix, so that $\mathbf{M}$ represents a lossless beam-combiner for which each column vector is of unit norm. As mentioned in appendix \ref{sec_normalization}, this condition on the norm is necessary (but not sufficient) to ensure that the matrix represents a lossless beam combiner, and one solution for it is to have:  
        \begin{equation}
            \begin{cases}\label{eq_normalization}
                a_l = \frac{1}{\sqrt{n_a}} & \text{for the bright output}\\
                a_l = \frac{1}{\sqrt{n_a}} \sqrt{\frac{n_a-1}{n_{null}}} & \text{for the nulled outputs}
            \end{cases}
        \end{equation}{}
        where $n_{null}$ is the number of nulled outputs. Normalization is not mandatory to study the qualitative properties of the combiner, but it is necessary to study their throughput in a quantitative manner and their practical implementation.\par
        
        The matrix $\mathbf{M}$ obtained with Eq. (\ref{eq_generative}), represents a combiner for which pairs of complex conjugate nulls can be subtracted to build the kernel nulls that are the focus of this work.

    \subsection{Information redundancy}\label{sec_info_redundancy}
        As shown by Eq. (\ref{eq_nmax}) and Table \ref{tab_nuller_growth}, the number of nulled outputs that would result from a strict application of these blueprint rules rapidly grows as the factorial of the number of inputs. However, for numbers of apertures larger than four, although all the nulls produced with the presented scheme are distinct, some of them do not carry new information on the target, as their response function to off-axis signal is a linear combination of the response function of other nulls.\par
        
        Here, we analyze this property empirically by examining the response maps (analogous to Fig. 5 and 7 of \cite{Martinache2018}) and assembling them as vectors of a set of nulled outputs and a set of kernel outputs. The ranks of these sets provide the number of independent observables produced by the combiner. Although we were not able to link this property to particular traits of the combinations, the largest number of independent kernel nulls obtainable by a given non-redundant array of apertures was always the same as the number of independent closure-phases, which is in agreement with the expectations set by \cite{Martinache2018}. For any non-redundant array of apertures, this number is:
        \begin{equation}\label{eq_na-2}
            n_{kn}(n_a) = \binom{n_a}{2} - (n_a - 1) =\frac{(n_a-1)(n_a-2)}{2}.
        \end{equation}
        The underlying relationship between the baselines and our new observables is non-trivial but will be assumed to hold for any non-redundant array. For redundant arrays, this number decreases. We will call complete a nuller that provides the aforementioned maximum number $n_{kn}$ of independent observables. The number of independent nulls in the full set is $n_{indep.} = 2\times n_{kn}$. These results obtained empirically for up to seven inputs are shown in Table \ref{tab_nuller_growth}, along with their expected progression for larger numbers of inputs.\par

        \begin{table}[]
            \centering
            \begin{tabular}{c|c|c|c}
                Inputs & Distinct nulls & Indep. nulls & Kernel nulls\\
                $n_a$ & $n_{max}$ & $n_{indep.}$ & $n_{kn}$ \\
                \hline
                3 & 2 & 2 & 1 \\
                4 & 6 & 6 & 3 \\
                5 & 24 & 12 & 6 \\
                6 & 120 & 20 & 10 \\
                7 & 720 & 30 & 15 \\
            \end{tabular}
            \caption{Growth of kernel-nuller combiners with the number of apertures.}
            \label{tab_nuller_growth}
        \end{table}{}
        
        As seen in Eq. (\ref{eq_normalization}), an increase in the number of nulled rows decreases the normalization coefficients $a_i$, as in practice fewer splittings of the input light are necessary to obtain the fewer combinations. Our goal may therefore be to construct complete combiners using the minimum number of nulled combinations from the full matrix $\mathbf{M}$, with the intent of increasing its throughput. Expecting this number to be twice the number of kernel nulls (if we consider only pairwise kernel nulls), this produces a very large number $n_{crops}(n_a)$ of possible combinations:
        \begin{equation}\label{eq_big_permutation}
            n_{crops}(n_a) = \binom{(n_a-1)!}{(n_a-1)(n_a-2)}
        \end{equation}
        Only for the cases of three and four inputs is the solution unique ($n_{crops}(n_a)=1$), and all null rows must be kept. For more inputs, this number grows rapidly. Although a large fraction of them are complete, fewer satisfy the conditions detailed in appendix \ref{sec_normalization} for conservation of energy.\par
        The following characteristics are shared by the three- and four-input combiners, as well as all the lossless realizations of the cropped five-input combiner:
        \begin{itemize}
            \item All nulls appear in conjugate (or enantiomorph) pairs, which implies that robust observables can be constructed by subtraction.
            \item Each phasor appears in each column the same number of times (except for the one of phase zero which serves as the reference). Equation (\ref{eq_na-2}) implies that for a combiner producing $n_o = 2n_{kn}$ nulls, each phasor is used $(n_a-2)$ times for each input.
        \end{itemize}
        Enforcing these characteristics as rules for reducing the number of outputs has helped us to identify lossless realizations of the cropped six-input combiner outlined in Sect. \ref{sec_6T_combiner} by reducing the parameter space. The process leading to a valid solution remains one of trial and error: chosen randomly among all the possible arrangements that respect the aforementioned characteristics, a solution is eventually only accepted when the corresponding combiner is both lossless (cf. appendix \ref{sec_normalization}) and produces a complete set of kernel nulls.\par

\section{Examples of combiners}
    
    \subsection{Three-input kernel nuller}\label{sec_3Tcombiner}
\begin{figure}
            \centering
            \includegraphics[width=0.45\textwidth]{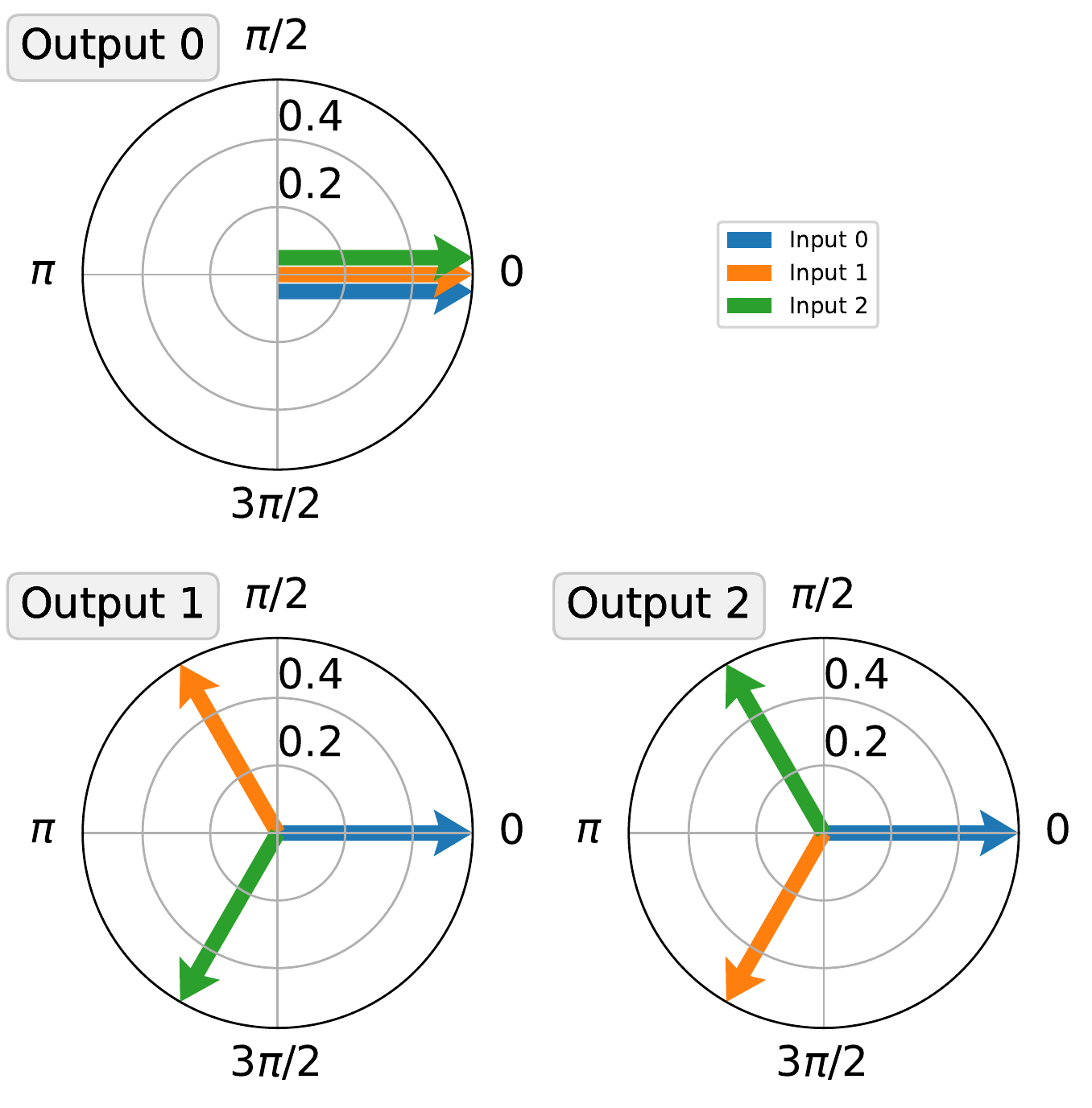}
            \caption{CMP for a three-input kernel nuller of Eq. (\ref{eq_3T_matrix}). The first row corresponds to the bright channel with the overlapping phasors staggered for readability. The two nulled outputs are complex conjugates of one-another and will form a kernel null.}
            \label{fig_3x3}
        \end{figure}

        The simplest practical example of this architecture appears for the combination of three inputs. Here, the algorithm results in the formation of two enantiomorph nulled outputs. Those two outputs will, by subtraction, produce one robust observable. The resulting combiner matrix writes:
        \begin{equation}\label{eq_3T_matrix}
            \mathbf{M}_{3} = 
            \frac{1}{\sqrt{3}}\left[\begin{matrix}1 & 1 & 1 &\\
                        1 &
                        e^{\frac{2 j \pi}{3}} & e^{\frac{4 j \pi}{3}}\\
                        1 & e^{\frac{4 j \pi}{3}} & e^{\frac{2 j \pi}{3}}\end{matrix}\right].
        \end{equation}
        The combinations offered by this matrix are illustrated in Fig. \ref{fig_3x3}.\par
        
\begin{figure}[H]
            \centering
            \includegraphics[width=0.45\textwidth]{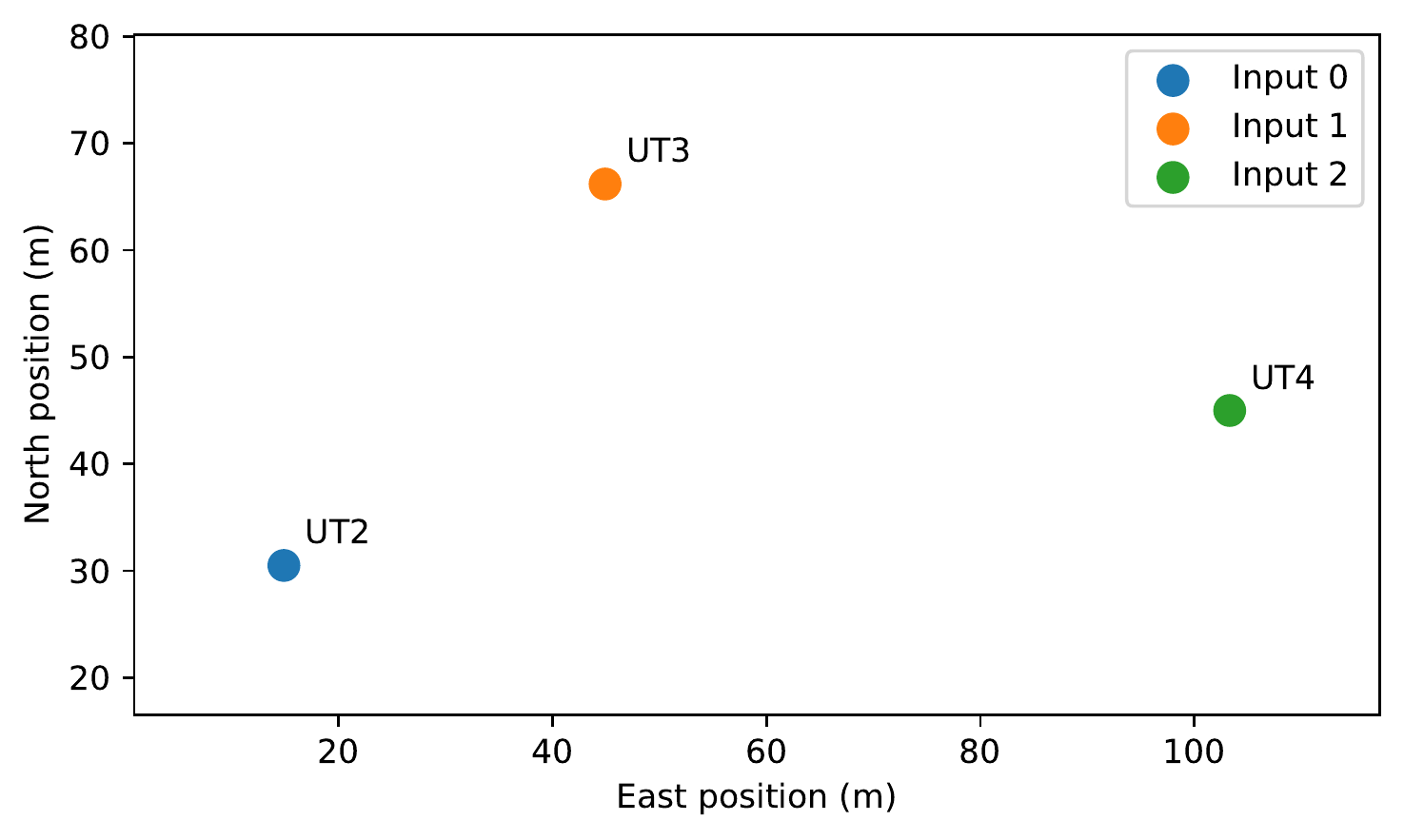}\par
            \includegraphics[width=0.40\textwidth]{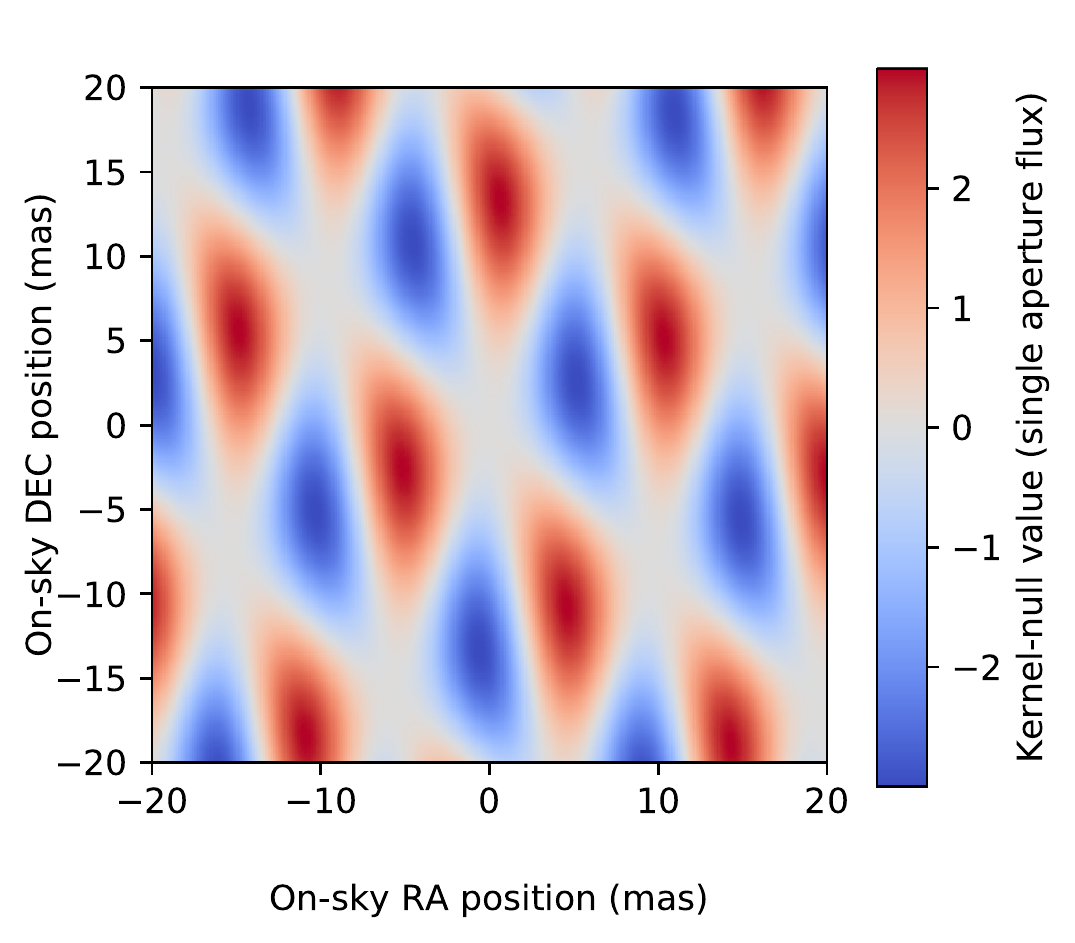}
            \caption{Top: the three telescope configuration picked for the example and corresponding to the position of three of the VLTI UTs. Bottom: the value of the kernel null as a function of the relative position of a source of unit contrast at $3.6\mu m$, normalized to the throughput of one aperture. The map is relevant for a target observed at Zenith and would evolve with the projected aperture map. Note the antisymmertric nature of the response, as demonstrated in Sect. \ref{sec_odd_response}}
            \label{fig_3T_maps}
        \end{figure}

        As an example, we built a response map of the robust observable produced by this combiner fed by three of the VLTI \citep{vonderLuhe1997} unit telescopes (UTs) observing at zenith. Figure \ref{fig_3T_maps} shows the values of the kernel-null observable represented as a two-dimensional function of the relative position of a source normalized by the flux of one aperture. While simpler than the one provided in Fig. 7 of \cite{Martinache2018} for the four-input combiner, this pattern retains the same antisymmetric property.\par
        
        %In this case, a trivial solution for the kernel matrix is the flat matrix  $\mathbf{K_3} = [1, -1]$ that performs the difference of the two nulled outputs intensities.\par
        
        Assuming the practical implementation of the combiner itself can be manufactured either with bulk or integrated optics, this configuration would allow the production of robust high-contrast observables with the least amount of infrastructure. Drawing a parallel between this type of combinations and closure triangles used for closure phases is tempting but misleading. Here, as the combination must be done optically rather than in post-processing, kernel nulling does not scale in the same way. The advantages and drawbacks of using these simple combiners as building blocks is briefly discussed in Sect. \ref{sec_modular}.\par

    \subsection{Six-input kernel nuller}\label{sec_6T_combiner}
        We also outline a solution for a kernel-nulling recombiner for six telescopes that could, for example, be used at the focus of the CHARA array. The initial algorithm produces a combiner matrix $\mathbf{M}_6$ with $121$ rows with redundancy in the off-axis response. It is cropped to $\mathbf{M}_6'$ using the guidelines offered in Sect. \ref{sec_info_redundancy} to reduce it to the minimum of $21$ rows while making sure the number of independent kernel nulls $n_{kn}$ is preserved. Furthermore, by enforcing the properties outlined in Appendix \ref{sec_normalization}, we make sure that $\mathbf{M}_6'$ remains the matrix of a lossless beam combiner.\par
        
\begin{figure}
            \centering
            \includegraphics[width=0.45\textwidth]{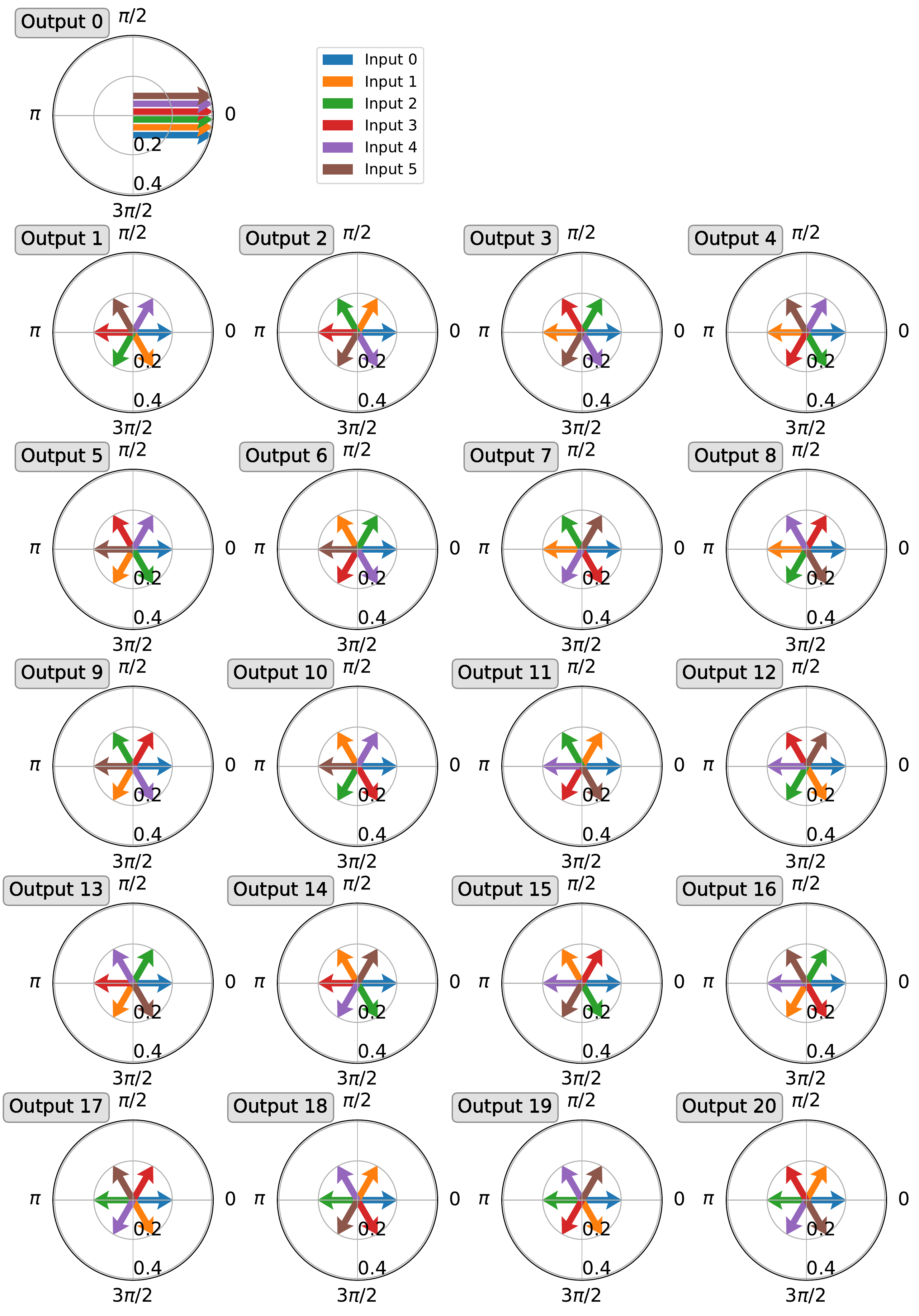}
            \caption{Representation of the 20 nulled outputs of a six input beam combiner proposed in Eq. (\ref{eq_6T_matrix}). The conjugate pairs that form the 10 kernel nulls are represented side-by-side.}
            \label{fig_6T_combiner}
        \end{figure}

        The matrix describing this 6-input combiner writes:
        \begin{equation}\label{eq_6T_matrix}
        \begin{split}
            \mathbf{M}_6' =
            \frac{1}{\sqrt{6}}\frac{1}{\sqrt{4}}\left[\begin{matrix}2 & 2 & 2 & 2 & 2 & 2\\1 & e^{\frac{5 j \pi}{3}} & e^{\frac{4 j \pi}{3}} & -1 & e^{\frac{j \pi}{3}} & e^{\frac{2 j \pi}{3}}\\1 & e^{\frac{j \pi}{3}} & e^{\frac{2 j \pi}{3}} & -1 & e^{\frac{5 j \pi}{3}} & e^{\frac{4 j \pi}{3}}\\1 & -1 & e^{\frac{j \pi}{3}} & e^{\frac{2 j \pi}{3}} & e^{\frac{5 j \pi}{3}} & e^{\frac{4 j \pi}{3}}\\1 & -1 & e^{\frac{5 j \pi}{3}} & e^{\frac{4 j \pi}{3}} & e^{\frac{j \pi}{3}} & e^{\frac{2 j \pi}{3}}\\1 & e^{\frac{4 j \pi}{3}} & e^{\frac{5 j \pi}{3}} & e^{\frac{2 j \pi}{3}} & e^{\frac{j \pi}{3}} & -1\\1 & e^{\frac{2 j \pi}{3}} & e^{\frac{j \pi}{3}} & e^{\frac{4 j \pi}{3}} & e^{\frac{5 j \pi}{3}} & -1\\1 & -1 & e^{\frac{2 j \pi}{3}} & e^{\frac{5 j \pi}{3}} & e^{\frac{4 j \pi}{3}} & e^{\frac{j \pi}{3}}\\1 & -1 & e^{\frac{4 j \pi}{3}} & e^{\frac{j \pi}{3}} & e^{\frac{2 j \pi}{3}} & e^{\frac{5 j \pi}{3}}\\1 & e^{\frac{4 j \pi}{3}} & e^{\frac{2 j \pi}{3}} & e^{\frac{j \pi}{3}} & e^{\frac{5 j \pi}{3}} & -1\\1 & e^{\frac{2 j \pi}{3}} & e^{\frac{4 j \pi}{3}} & e^{\frac{5 j \pi}{3}} & e^{\frac{j \pi}{3}} & -1\\1 & e^{\frac{j \pi}{3}} & e^{\frac{2 j \pi}{3}} & e^{\frac{4 j \pi}{3}} & -1 & e^{\frac{5 j \pi}{3}}\\1 & e^{\frac{5 j \pi}{3}} & e^{\frac{4 j \pi}{3}} & e^{\frac{2 j \pi}{3}} & -1 & e^{\frac{j \pi}{3}}\\1 & e^{\frac{4 j \pi}{3}} & e^{\frac{j \pi}{3}} & -1 & e^{\frac{2 j \pi}{3}} & e^{\frac{5 j \pi}{3}}\\1 & e^{\frac{2 j \pi}{3}} & e^{\frac{5 j \pi}{3}} & -1 & e^{\frac{4 j \pi}{3}} & e^{\frac{j \pi}{3}}\\1 & e^{\frac{2 j \pi}{3}} & e^{\frac{5 j \pi}{3}} & e^{\frac{j \pi}{3}} & -1 & e^{\frac{4 j \pi}{3}}\\1 & e^{\frac{4 j \pi}{3}} & e^{\frac{j \pi}{3}} & e^{\frac{5 j \pi}{3}} & -1 & e^{\frac{2 j \pi}{3}}\\1 & e^{\frac{5 j \pi}{3}} & -1 & e^{\frac{j \pi}{3}} & e^{\frac{4 j \pi}{3}} & e^{\frac{2 j \pi}{3}}\\1 & e^{\frac{j \pi}{3}} & -1 & e^{\frac{5 j \pi}{3}} & e^{\frac{2 j \pi}{3}} & e^{\frac{4 j \pi}{3}}\\1 & e^{\frac{5 j \pi}{3}} & -1 & e^{\frac{4 j \pi}{3}} & e^{\frac{2 j \pi}{3}} & e^{\frac{j \pi}{3}}\\1 & e^{\frac{j \pi}{3}} & -1 & e^{\frac{2 j \pi}{3}} & e^{\frac{4 j \pi}{3}} & e^{\frac{5 j \pi}{3}}\end{matrix}\right] .
        \end{split}
        \end{equation}%6T recioe 0
        This combiner offers a total of 20 independent nulls, and 10 independent kernel nulls. The corresponding CMP is showed in Fig. \ref{fig_6T_combiner} and highlights how each aperture contributes to all of the outputs.  \par
        
\begin{figure}
            \centering
            \includegraphics[width=0.3\textwidth]{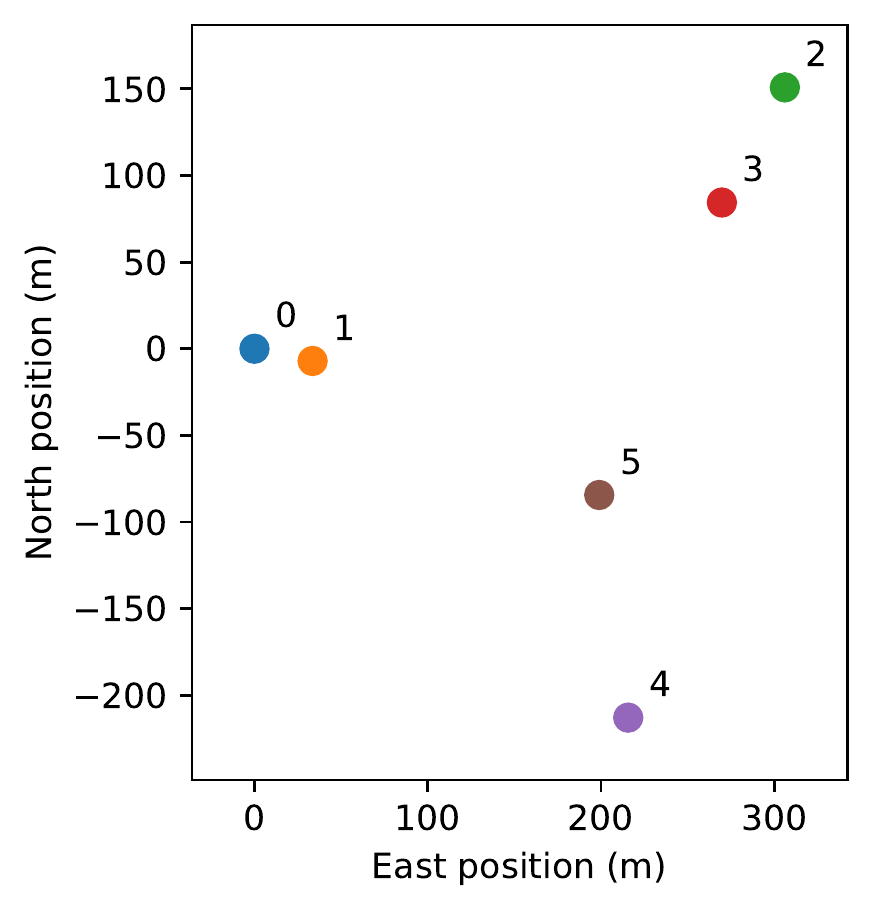}\par
            \includegraphics[width=0.45\textwidth]{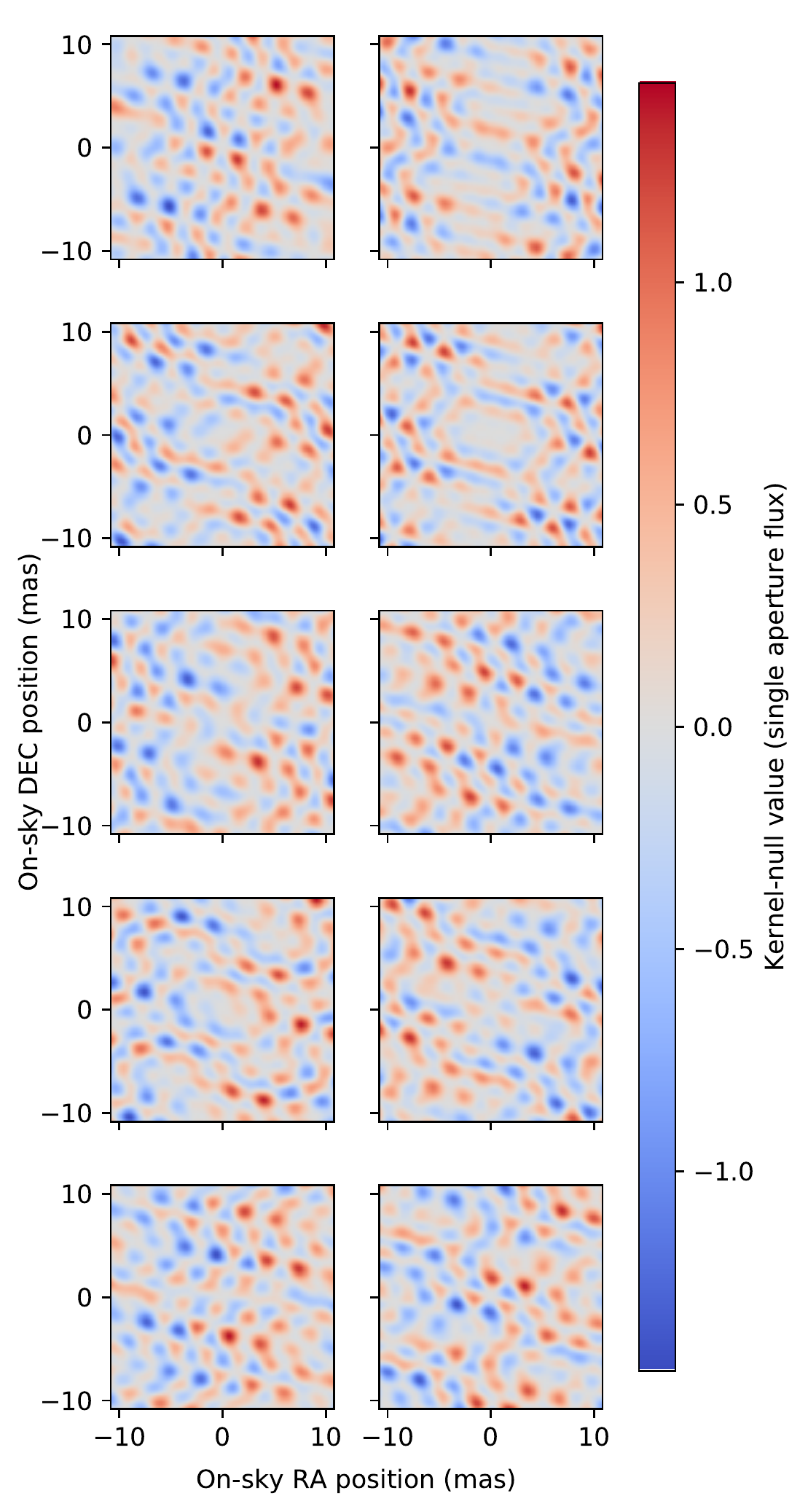}
            \caption{Top: the six telescope configuration for the CHARA array used as example. Bottom: the value of all 10 kernel nulls as a function of the relative position of a source at the wavelength $3.6\mu m$ observed at zenith. The transmission is normalized by the flux of a single aperture. Again, each map remains antisymmetric.}
            \label{fig_6T_maps}
        \end{figure}

        To illustrate the astrophysical information gathered by the larger number of kernel nulls, we construct response maps of the kernel-null observables. These plots, shown in Fig. \ref{fig_6T_maps} display the response of each of the observables for the combiner being fed by the CHARA array observing a target at zenith in the $3.6\mu m$ wavelength. The patterns reflects the richness of the uv coverage provided by an array like CHARA, and the fact that each output uses information collected by every telescope. As a consequence, each map covers the field of view differently, and brings new constraint on the properties of the astrophysical scene observed.\par
        
\section{Discussion}
    No active long-baseline optical interferometer currently provides more than six sub-apertures. However, the masking of monolithic apertures to produce interferometric arrays is an established practice \citep{Tuthill2010a, Jovanovic2012} that may be used in conjunction to nulling interferometry \citep{Norris2019}. Therefore, the use of even larger combiners may prove to be a viable alternatives to small inner working angle coronagraphs \citep{Guyon2006a}. Their robustness to small aberrations might provide unprecedented contrast performance in the $1-3 \lambda /D$ regime in the near infrared.\par
    
    \subsection{Multiplexing nullers} \label{sec_modular}
    
        Instead of building an all-in-one combiner, which may be difficult to construct for a large number of apertures, an alternative approach would be to multiplex several independent nullers, that each recombines a smaller number of apertures. For example, instead of a six-input nuller producing 20 nulled outputs, one conservative option would be to use two three-input kernel nullers, identical to the one presented in Sect. \ref{sec_3Tcombiner}, side by side producing four nulled outputs.\par
        
        While the latter of these two options results in a reassuring higher throughput per output, it can only produce distinct robust observables, where the $\mathbf{M}_6'$ combiners offers 10. Moreover, this multiplexed option also results in two bright channels, where some of the off-axis light is also lost, further reducing the overall efficiency of the combiner. In between these two extreme scenarios, intermediate solutions can be imagined to alleviate some of their risks and deficiencies, with a modular design multiplexing the nullers much like a number of ABCD combiners are multiplexed inside the beam combiner of VLTI/GRAVITY.
        
        It was already argued by \cite{Guyon2013} that efficient nulling solutions concentrate the most starlight into the smallest number of outputs, which favors the all-in-one combiner over the multiplexed versions. If manufacturability or operational constraints were to prevent the deployment of an all-in-one combiner at the focus of a specific observatory, one way to alleviate this inefficiency could be to recombine the light from the multiple bright outputs, into an additional nulling stage so as to extract additional useful observables. This type of architecture, in part inspired by the hierarchical fringe tracker idea of \cite{Petrov2014}, might prove a necessary compromise to the implementation of a kernel nuller at the focus of a very long baseline observing facility such as the envisioned Planet Formation Imager \citep{Monnier2016}, for which a distributed hierarchical recombination mode seems particularly apt.
        
    \subsection{Evolution of robustness}
        In addition to trade-off considerations between the total number of observables and the throughput efficiency of the available options, one must also consider whether the number of inputs has an impact on the phase-noise rejection performance of a kernel nuller.\par
        
\begin{figure}
            \centering
            \includegraphics[width=0.48\textwidth]{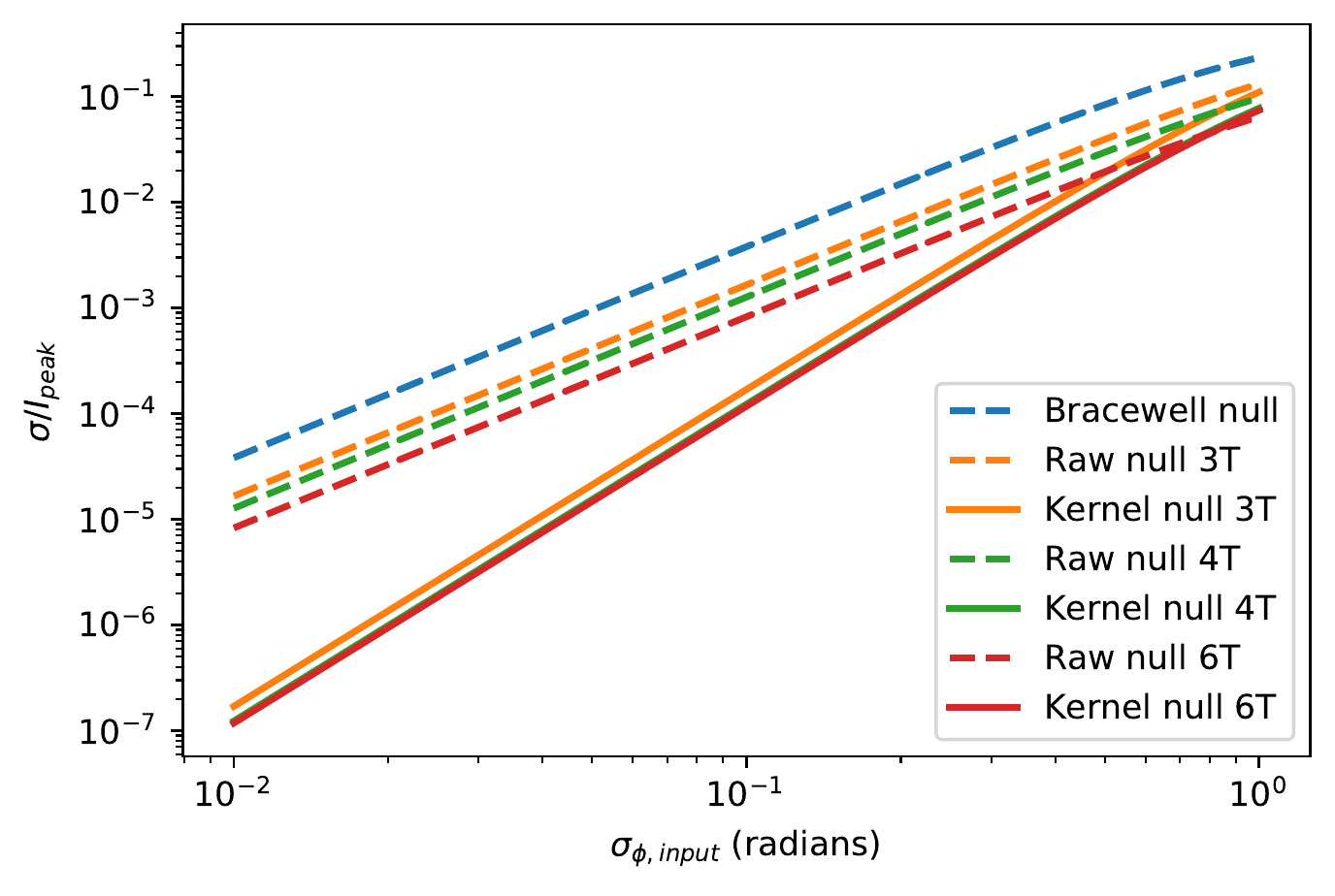}
            \caption{Propagation of phase noise from the input to the nulled intensities for different kernel-nulling architectures. Values are normalized by the peak transmission of an off-axis signal. The raw nulls (dashed lines) are compared with their corresponding kernel-null observable (solid lines), showing the suppression of 2nd order phase noise (by a few orders of magnitude) for small input phase error. This effect decreases as the input phase errors depart from the small phase approximation. The behavior of the Bracewell nuller is shown in dashed blue for reference.}
            \label{fig_plot_comparison}
        \end{figure}

        To evaluate this risk, we trace the evolution of the noise affecting the outputs and their kernels as a function of the amount of phase noise affecting the inputs. We do this for the 3T, 4T and the 6T designs described in the previous sections. For simplicity, this study assumes that the phase noise affecting all inputs is Gaussian, non-correlated, and characterized by a single rms value equally affecting all inputs. Following a Monte Carlo approach, random realizations of input piston errors are drawn, propagated through the different combiner matrices, and the standard deviation of the output intensities are evaluated. Figure \ref{fig_plot_comparison} thus shows the evolution of the standard deviation of the output intensities of the nulls and of the kernel nulls of the different architecture normalized by the peak null intensity $I_{peak}$ from their response map. This therefore constitutes a noise-to-signal ratio of sorts. The simple Bracewell nuller is also added to this study for comparison, as modelled by the combiner matrix:
        \begin{equation}
            \mathbf{M}_B = \left[\begin{matrix}1 & 1\\
                                            - j & j\end{matrix}\right] .
        \end{equation}
        \par

        {}
        
        This plot shows that the larger kernel-nulling combiners provide a rejection of the phase noise that is very similar to the smaller ones, if not slightly better. The improvement on the raw observables may be credited to a manifestation of the central limit theorem affecting the distribution of the sum of a larger number of complex intensities. Further interpretation of this plot must be undertaken with caution. Indeed, while the distribution of kernel nulls under such conditions is close to Gaussian \citep{Martinache2018}, the distribution of null intensities is not \citep{Hanot2011} and is therefore poorly described by its standard deviation. While a full performance comparison of the different designs lies outside the scope of the present paper and would include the coupled effects of phase and amplitude fluctuations \citep{Lay2004}, these elements already indicate that kernel nullers recombining a large number of sub-apertures are intrinsically at least as robust to phase noise as their smaller, simpler counterparts. This is an encouraging prospect for single telescope applications of the kernel nuller for which a potentially large number of sub-apertures can be used.

\section{Conclusions}
    In this work, we offer a new description of the kernel-nuller design introduced by \cite{Martinache2018}. This is done by introducing a new graphical representation of the complex matrix that models the nuller and the transformations it operates on the input electric field. Combined with an analytical description of the outputs used to form a kernel, these representations explain the origin of a kernel nuller's main properties: their intrinsic robustness to small input piston and amplitude fluctuations, and their sensitivity to asymmetric features of the observed scene. We incidentally show that the same outputs can also be summed so as to fall back on the original outputs of an all-in-one $\mathbf{N}_4$ nuller stage, while not robust to perturbations, can nevertheless provide further astrophysical information.\par
    
    Our graphical and analytical representations help devise a systematic way to build a kernel nuller as a combiner featuring pairs of channels that are enantiomorph in the complex plane. It is this feature that makes two channels equally sensitive to perturbations although they respond differently to the presence of an off-axis structure. This approach allow us to design kernel nullers for an arbitrary number of apertures, which we here apply to three- and six-aperture arrays.\par
    
    We discuss the possibility of simplifying kernel nullers that grow in complexity when they recombine a larger number of input beams, for instance, using distinct nullers operating in parallel over a subset of input beams. For a given total number of inputs, a global architecture, giving access to a larger number of high-contrast observables is more efficient and offers the means to explore and characterize complex astrophysical scenes. For the same number of inputs, we can also note that the total number of outputs for a kernel nuller (exactly twice the number of theoretical closure-phases) is in fact less than that of non-nulling combiners designed to measure the complex visibility of all baselines. Integrated optical circuits in particular already enable the implementation of such complicated designs in small and stable packages, and are a very promising avenue for the construction of these larger kernel-nulling combiners.\par
    
    While no existing long baseline optical interferometric facility currently offers the simultaneous combination of more than six apertures, a kernel nuller sampling the pupil of a single telescope could prove to be a valuable complement to a coronagraph, producing high contrast observations near one resolution element that would be insensitive to the small but ever present adaptive optics residuals. The evaluation of performance in practical implementations including the contribution of coupled phase an amplitude contributions and the consideration of relevant science cases will be the topic of future theoretical and experimental work.

\begin{acknowledgements}
    We thank Alban Ceau and Coline Lopez for their suggestions to improve the manuscript. 
    KERNEL has received funding from the European Research Council (ERC) under the European Union's Horizon 2020 research and innovation program (grant agreement CoG - 683029).
\end{acknowledgements}

\bibliographystyle{aa}
\bibliography{kernel_nulling}

\begin{appendix}
\section{Lossless combiners}\label{sec_normalization}
    As emphasized by \cite{Loudon2000} in section 3.2, the conservation of energy in a single beamsplitter cube imply that its matrix is unitary. This property can be generalized to larger combiners. The required condition is that the sum of intensities of the inputs be the same as the sum of intensities at the outputs. As the output intensities are gathered in the vector $\mathbf{x}$, this sum also writes:
    \begin{equation}
        \sum_{i=0}^{n_{outputs}} |x_i|^2 = \mathbf{x}^H \mathbf{x},
    \end{equation}{}
    with $^H$ designating the Hermitian adjoint (conjugate transpose) operator. Based on Eq. (\ref{eq_xMz}) this sum writes:
    \begin{equation}
        \mathbf{x}^H \mathbf{x} = (\mathbf{M}\mathbf{z})^H \mathbf{M}\mathbf{z},
    \end{equation}{}
    which develops as follows:
    \begin{equation}
        \mathbf{x}^H \mathbf{x} = \mathbf{z}^H\mathbf{M}^H \mathbf{M}\mathbf{z}.
    \end{equation}{}
    As a consequence we obtain:
    \begin{equation}
        \forall \mathbf{z} \in \mathbb{C}^{n_a} \sum_{i=0}^{n_{outputs}} |x_i|^2 = \sum_{i=0}^{n_a} |z_i|^2 
        \iff
        \mathbf{M}^H \mathbf{M} = \mathbf{I}.
    \end{equation}{}
    As a consequence, the following propositions are equivalent:
    \begin{itemize}
        \item $\mathbf{M}$ is the matrix of a lossless beam combiner.
        \item $\mathbf{M}$ is semi-unitary on the left.
        \item $\mathbf{M}^H$ is the left inverse of $\mathbf{M}$.
        \item The columns of $\mathbf{M}$ are orthonormal.
        \item All the singular values of $\mathbf{M}$ are equal to one.
    \end{itemize}

\end{appendix}{}
%\tableofcontents

\end{document}